\documentstyle[11pt]{article}

\advance \topmargin by -\headheight
\advance \topmargin by -\headsep

\evensidemargin \oddsidemargin
\newcommand{\be}{\begin{equation}}
\newcommand{\ee}{\end{equation}}
\newcommand{\bea}{\begin{eqnarray}}
\newcommand{\eea}{\end{eqnarray}}
\newcommand{\nen}{\nonumber \\ \relax}

\newcommand{\forcepar}{{\hskip 10pt\vskip -15pt}}
\newfont{\headfont}{cmbx10 scaled 1440}
\newfont{\namefont}{cmr10}
\newfont{\initialfont}{cmr10 scaled 1200}
\newfont{\addfont}{cmti7 scaled 1440}
\newfont{\boldmathfont}{cmbx10}
\newfont{\figfont}{cmr7 scaled 1200}

\newcommand{\seq}{\ =\ }
\newcommand{\pls}{\ +\ }
\newcommand{\mi}{\ -\ }

\newcommand{\half}{\frac{1}{2}}
\newcommand{\inv}[1]{\frac{1}{#1}}

\newcommand{\ca}{{\cal A}}

\newcommand{\cf}{{\cal F}}

\newcommand{\cm}{{\cal M}}

\newcommand{\cs}{{\cal S}}
\newcommand{\ct}{{\cal T}}

\newcommand{\IR}{{I \kern -0.4em R}}
\newcommand{\IC}{{I \kern -0.65em C}}
\newcommand{\ZZ}{{Z \kern -0.8em Z}}

\newcommand{\np}[1]{{\it Nucl. Phys.} {\bf B#1}}
\newcommand{\cmp}[1]{{\it Commun. Math. Phys.} {\bf #1}}

\def\ca{{\cal A}}

\def\cf{{\cal F}}

\def\cm{{\cal M}}

\def\cs{{\cal S}}
\def\ct{{\cal T}}

\def\hf{\frac{1}{2}}

\def\c{\chi}

\def\e{\epsilon}
\def\f{\phi}

\def\h{\eta}

\def\j{\psi}

\def\l{\lambda}

\def\n{\nu}

\def\p{\pi}

\def\J{\Psi}

\def\S{\Sigma}

\newcommand{\IZ}{{Z \kern -0.67em Z}}
\let\nopictures=Y
\newfont{\headfontb}{cmbx10 scaled 1728}
\begin{document}
\begin{titlepage}
\renewcommand{\thefootnote}{\fnsymbol{footnote}}
\begin{center}
{\headfontb The Monopole Equations\\ ~in Topological Yang-Mills}\footnote{This
work is supported in part by funds
provided by the
U. S. Department of Energy (D.O.E.) under cooperative agreement
\#DE-FC02-94ER40818.}

\end{center}
\vskip 0.3truein
\begin{center}
{
{\Large R}{OGER} {\Large B}{ROOKS\footnote{Email: rog@ctpup.mit.edu}}
{AND}
{\Large A}{RTHUR} {\Large L}{UE\footnote{Email: ithron@mit.edu}}
}
\end{center}
\begin{center}
{\addfont{Center for Theoretical Physics,}}\\
{\addfont{Laboratory for Nuclear Science}}\\
{\addfont{and Department of Physics,}}\\
{\addfont{Massachusetts Institute of Technology}}\\
{\addfont{Cambridge, Massachusetts 02139 U.S.A.}}
\end{center}
\vskip 0.5truein
\begin{center}
\bf ABSTRACT
\end{center}
We twist the monopole equations of Seiberg and Witten and show how these
equations are realized in topological Yang-Mills theory.  A Floer derivative
and a Morse functional are found and are used to construct a unitary
transformation between the usual Floer cohomologies and those of the monopole
equations.  Furthermore, these equations are seen to reside in the vanishing
self-dual curvature condition of an $OSp(1|2)$-bundle.  Alternatively, they may
be seen arising directly from a vanishing self-dual curvature condition on an
$SU(2)$-bundle in which the fermions are realized as spanning the tangent space
for a specific background.
\vskip 7truecm
\leftline{CTP \# 2397  \hfill December 1994}
\smallskip
\leftline{hep-th/9412206}

\end{titlepage}
\setcounter{footnote}{0}
\section{Introduction}
\forcepar
In this note, we will demonstrate how the monopole equations  of ref.
\cite{W(mon)} for an abelian connection $A$ and $SU(2)$  doublet fermions arise
in topological Yang-Mills gauge (TYM) theory \cite{W(TYM)} and in Floer theory
\cite{Fl} in particular.  As we will see the process is remarkably simple.
Along the way, we will develop a quantum mechanical system whose ground states
have support on the fields which satisfy the twisted monopole equations. What
is more, we will find that the inner products of representatives of the
cohomology groups so constructed, are formally equal to the  Donaldson
invariants \cite{Don}.

Consider the twisted version of the monopole equations.  Let $\ct$ denote the
twisting map, $S^\pm$ the right/left spin bundles over a four dimensional
manifold, $X$, $\Lambda^p$ the bundle of $p$-forms and $\Lambda_O^p$ the bundle
of $p$-forms with odd Grassmann parity.  Then for a spinor  which in addition
to being a section of $S^+$ is also a doublet of a rigid $SU(2)$ (denoted by
$\cs$) so that $M=\sigma(S^+\otimes\cs)$, we have
$\ct: \sigma(S^+\otimes\cs)\to \Lambda_O^1$;
likewise\footnote{$P_\pm(\Lambda^2)=\Lambda^{2\,\pm}$ is the projection to
(anti-)self-dual two-forms.}, $\ct: \sigma(S^-\otimes\cs)\to
\Lambda_O^0\oplus\Lambda_O^{2\,+}$.  With this as background,  the equations we
are interested in are
\be
P_+(D\psi)\seq0\ \ ,\qquad D^*\psi\seq0\ \ ,\qquad F^+\seq
iP_+(\bar\psi\wedge\psi)\ \ ,\label{E_TSW}\ee
where $\psi=\sigma(\Lambda_O^1\otimes L)$, $D$ is the covariant exterior
derivative given by the connection $A$ on $L$, $P_+=\half(I+*)$ is the
self-dual projector and the ``bar" denotes complex conjugation.  In the
notation of ref. \cite{W(mon)}, the elliptic complex on which these  twisted
monopole equations are realized form the exact sequence
\be
0~\longrightarrow~ \Lambda^0~\stackrel{s}{\longrightarrow}~
\Lambda^1\oplus(\Lambda_O^1\otimes L)~\stackrel{t}{\longrightarrow}~
\Lambda^{2\,+}\oplus(\Lambda_O^0\oplus\Lambda_O^{2\,+}) ~\longrightarrow~0\ \
.\ee
This complex defines the arena in which we will work in this paper.
These equations stand on their own irrespective of our discussion in the
previous paragraph; in particular, the value of $w_2(X)$.  However, they may be
viewed as arising from the twisting of the N=2 versions of the monopole
equations on spin manifolds.  It should be noted that, unlike TYM, the first
equation cannot be obtained by smoothly varying the connection in the third
equation.

In the next section we will see how the twisted monopole equations
(\ref{E_TSW}) arise by an reduction of a class of zero-action solutions of TYM
to a $U(1)$ subgroup of the gauge group. Our intention is not to perform a
duality transformation on or any addition (such as twisted N=2 hypermultiplets)
to TYM as we would like to see these equations directly in the field space of
the quantum field theory for the Donaldson invariants.  In this way, we hope to
be able to shed light on the connection between the Seiberg-Witten invariants
\cite{{W(mon)},W(SYM),SW,{KM1}} and those of Donaldson \cite{Don}.  A step in
this direction will be made in section \ref{S_Floer}.  Some other directions
spawned by this approach will be discussed in section \ref{S_Novel}.  Our
conclusions may be found in section \ref{S_Conc}.

\section{In Topological Yang-Mills}\label{S_TYM}
\forcepar
Let us focus on obtaining the twisted monopole equations as a special
minimum action condition in topological Yang-Mills theory (TYM) on a
$G$-bundle.  For simplicity, we will take the structure group to be $SU(2)$.

First, we realize that the eqns. (\ref{E_TSW}) cannot arise by breaking the
gauge group to $U(1)$.  Although breaking the gauge group in TYM is possible by
adding\footnote{Note that $V(\phi)$ is
a Higgs potential for the BRST singlet field, which is bounded from below at
zero. Although this explicitly introduces a metric (in a volume preserving,
diffeomorphism invariant way), as long as $V$ is zero in the low-energy
effective field theory, we can ignore this effect.}
\be
S_m~\equiv~\{Q, \int_X Tr(\psi\wedge ^*D\phi\})\pls \int_X V(\phi)\ \ ,\ee
to the action for TYM, the equations cannot be obtained in the low-energy
effective theory as all fields are in the adjoint representation here.  It is
presumably possible to add matter to the TYM theory so that the fermions which
appear on the right-hand-side (via a current-current coupling in the effective
action) of eqn. (\ref{E_TSW}) were not in the adjoint  representation of the
gauge group.  However, a negative feature of that approach would be to take us
outside of the field space of TYM, thus making it difficult to realize the
connection with the Donaldson invariants.  Thus, we now resort to the explicit
breaking of the $SU(2)$ gauge group.

As an ansatz, let us look for field configurations for which the TYM action
\cite{W(TYM)} vanishes.    Write the $SU(2)$ generators as $J^a\equiv (J,
\bar{J},J^3)$, similarly\footnote{We will use the symbol $\ca^a$, with gauge
index $a$, to denote the $SU(2)$ connection; while we will use $A=\ca^3$ for
the $U(1)$ connection, as in the previous section.} for the sections of
$ad(G)$.  Take the following fields to vanish:  $\ca, \bar{\ca}, \l, \bar{\l},
\f^3, \c^3,$ $\h^3$.  We will in addition set $(\f,\bar{\f}) = (\n,\bar{\n})$,
where for now $\nu$ is a complex constant.  Note that the BRST transformations
in this field-restricted sector are $[Q',A] = \j^3$ and $\{Q',\j^a\} = 0$.

After integrating out $\l^3$, the TYM action becomes
\bea
	S &\seq& \int_X \left[\frac{1}{8}F^{3+}\wedge ^*F^{3+} \pls \frac{1}{8|\n|^2}
(\j\wedge\bar{\j})\wedge^*(\j\wedge\bar{\j}) \right.	\nen
& & \left. \frac{}{}\mi \bar{\c}\wedge ^*D\j \mi \c\wedge ^*D\bar{\j} \pls
\bar{\h} D^*\j
\mi \h D^*\bar{\j}\right]
\ \ .\eea
The first line in this expression may be written as

\be
	S_0 \seq \frac{1}{8}	\int_X  \left| F^{3+} \pls
\frac{1}{\n}P_+(\bar{\j}\wedge\j)
	\right|^2
\ \ ,\ee
so long as $\n$ is pure imaginary.  Then, invoking the $\c,\h$-equations of
motion, we see that the action is zero on non-trivial field equations which
satisfy the twisted monopole equations (\ref{E_TSW}) with $\j \rightarrow
|\n|^{-\hf}\j$.

Now that we have obtained the twisted monopole equations in Donaldson field
space, we might impose them as semi-classical conditions on the polynomial
invariants.  Indeed, the parameter $|\n|$ is best defined by the Donaldson
invariant

\be	\langle W_0 \rangle\seq \langle\hf\hbox{Tr}(\f^2)\rangle \seq
\frac{1}{4}|\n|^2
\ \ ,\ee
where the brackets mean the evaluation on these special field configurations.
For the map from $H_2(X)$ to $H^2(\cm)$, the observable, $\int_\S W_2$, becomes

\be
\langle\int_\S W_2\rangle\seq
\langle\int_\S \hbox{Tr} (\hf\j\wedge\j + \f F)\rangle \seq -2\p|\n| c_1(L)[\S]
\ \ ,
\ee
proportional to the first Chern class of the line bundle.

Having recovered the twisted monopole equations by hand from TYM, we now wonder
why they should exist in the latter theory in the first place.  Well, this is
where the ``by-hand" procedure we have just performed actually teaches us
something.  The first and third equations in (\ref{E_TSW}) are nothing but the
anti-self-dual condition in disguise.
To see this, consider starting off with the equation $\cf^+(\ca)=0$ for a
$SU(2)$ curvature with connection $\ca$.  Then write the connection as a
particular background $(A)$ plus a fluctuation $(\hat\psi)$ via the expressions
$\ca^3=A^3$, $\inv{\sqrt{2}}(\ca^1+ i\ca^2)=\hat\psi$ and
$\inv{\sqrt{2}}(\ca^1- i\ca^2)=\hat{\bar\psi}$; {\it i.e.}, $A^3$ does not
fluctuate while the other connection components do not have background parts.
Upon substituting these expressions into the $\cf^+=0$ equation, dropping the
hats on the $\psi$'s and changing their Grassmann parity we arrive at the first
and third twisted monopole equations.  Thus we see that it is not surprising
that we obtained them in TYM.

\section{In Floer Cohomology}\label{S_Floer}

\forcepar
We have seen how the twisted monopole equations appear as a minimum action
configuration in TYM.  We will now identify the analogous Floer cohomology
condition.  Then we will see that a unitary transformation exists which relates
the Floer and monopole cohomologies so constructed.   It is important that we
will be working in the phase space of the Floer theory.  As a point of
reference recall that given a closed, orientable $3$-manifold, $Y$, the Floer
cohomology operator is ($t$ is an arbitrary real parameter)
\be
Q_t\seq \int_Y  \psi_i{}^a(x)\left(\frac{\delta}{\delta \ca_i{}^a(x)} \pls \hf
t \epsilon^{ijk}\cf_{jk\, a}(x)\right)\ \ ,\ee
for which the representatives of the cohomology groups are the wavefunctionals,
$\Psi[\ca_i{}^a,\psi_i{}^a]$ which satisfy the condition $Q_t\Psi=
Q^\dagger_t\Psi = 0$, where $\j^\dagger = \bar\c$.

As before, let $\ca\equiv(\ca^\pm,A)$ be the connection on the $SU(2)$ bundle,
$G$, over $Y$ and take $\psi^a$ to be the components of a section of the bundle
$(\Lambda_O^1\otimes G)$.  Choose the Morse function to be (see also ref.
\cite{KM2})
\be
W'[A,\psi] \seq \inv{4\pi}\int_Y \left[A\wedge dA \pls\frac{}{}i 2\bar{\j}
\wedge D(A) \j \right] \ \ ,\ee
and based on $Q=\int_Y \psi_i{}^a\frac{\delta}{\delta \ca_i{}^a}$ define the
exterior derivative
 \bea
Q_t'&\seq&e^{-2\pi t W'[A,\j]}Q e^{2\pi t W'[A,\j]}\nen
&\seq&\int_Y \left[\psi_i{}^1\frac{\delta}{\delta \ca_i{}^1}\pls
\psi_i{}^2\frac{\delta}{\delta \ca_i{}^2}\right.\nen
&&\left.\pls
\psi_i{}^3(\frac{\delta}{\delta A_i} \pls \hf t \epsilon^{ijk}F_{jk}(A) \mi 2it
\epsilon^{ijk}(\bar{\psi}_j{}(x)\psi_k{}(x) ))
\, \right]
\ \ .\eea
Clearly, a solution of $Q_t'\Psi'[\ca_i{}^a,\psi_i{}^a]=0$ is any
$\Psi'[A,\psi]$ which has support only on equation (\ref{E_TSW}) written on
$X=Y\times\IR$,
\be
F_{0i}^+ \seq \epsilon_{ijk} \bar \psi^j\psi^k\ \ ,\ee
in the gauge $\ca_0{}^a=\psi_0{}^a=0$.

The hamiltonian whose vacuum states include solutions to the twisted monopole
equations will take the form $H' = \hf\{Q',Q'{}^\dagger\}$.  After some
straightforward algebra, one finds the new hamiltonian on the states
$\J'[A,\j]$ takes the form

\be
	H' \seq 2\int_Y (F^+_{0i} - \e_{ijk}\bar\j_j\j_k)^\dagger(F^+_{0i} -
\e_{imn}\bar\j_m\j_n)
\label{SWH}
\ee
Comparing this hamiltonian  to the Floer hamiltonian, we note
some interesting differences.  First, only the abelian component of the gauge
fields play a role.  Next, the fermionic partners appear in a fashion  which
does not preserve ghost number.  Note also that none of the fermions have
appropriate kinetic contributions.  All of these features are consistent with
the fact that the twisted monopole equations are simply re-writings of the
self-dual curvature condition.

The question remains how to extract the solution $\Psi'$ from the Floer
cohomology.  That is we seek a $W$  such that given a Floer representative
$\Psi$,
\be
\Psi[\ca_i{}^a,\psi_i{}^a] \seq e^{-2\pi tW[\ca_i{}^a,\psi_i{}^a]} \Psi'[A,\j]\
\ .\label{E_MAP}\ee
It is not hard to see that such a functional is given by
\be
W[\ca_i{}^a,\psi_i{}^a]~\equiv~ -~\inv{4\pi}\int_Y Tr( \ca d\ca + \frac{2}{3}
\ca^3)\pls W'[A,\j]\ \ ,\ee
Here, the first term in $W$ is recognized as the $SU(2)$ Chern-Simons action.
The virtue of the construction (\ref{E_MAP}) is that it allows us to conjecture
that
given $X\equiv X_l\cup_YX_r$, where $X_l$ and $X_r$ are manifolds whose
boundaries are diffeomorphic to $Y$ but have opposite orientation, the inner
products are related by
\be
\langle \Psi |\Psi\rangle\seq \langle \Psi' |\Psi'\rangle\ \ .\ee
It is reasonable to presume that the $\Psi'$ are representatives of a Floer
homology group but for spectral flows governed by the monopole equations and
are obtained from the Seiberg-Witten invariants via surgery.
In that case,  we conjecture that this equality will unlock the formal
relationship between the those invariants and the Donaldson polynomials.

\section{Other Directions}\label{S_Novel}
\forcepar
Apart from the obvious solutions to the usual Floer homology condition, namely
$\Psi$'s which have support only on  $\frac{\delta}{\delta A_i{}^a(x)} \pls \hf
t \epsilon^{ijk}\cf_{jk\, a}(x)=0~\leftrightarrow~ \cf_{0i\, a}^+=0$,  another
simple solution is evident.  If the condition
\be
\cf_{0i\, a}^+\seq\kappa f_{abc}\epsilon_{ijk}\psi^{j\, b}\chi^{k\, c}\ \ ,\ee
is met, then $Q_t= Q^\dagger_t = 0$ for any $\kappa$.  However, this solutions
is not compatible with equation (\ref{E_TSW}) due to the presence of the
structure constants and the fact that $\psi$ and $\chi$ are canonically
conjugate to each other.  Beyond this, our methodology in the last section may
be extended to construct other cohomologies.

Our procedure suggests another direction to explore.  As we discussed before,
we do not want to add topological matter to TYM in order to obtain the monopole
equations as this would spoil the direct connection with the self-dual
curvatures of Donaldson theory.  Now, in principle, TYM exists for an arbitrary
structure group.  With this in mind, we are led to introduce a group which has
both bosonic and fermionic generators; i.e., a supergroup.  For simplicity, let
us take the group to be $OSp(1|2)$ with graded commutators:
\bea
[J_a,J_b]&\seq& \epsilon_{abc}J^c\ \ ,\nen
[J_a,Q_\alpha]&\seq& -i\half (\gamma_a)_\alpha{}^\beta Q_\beta\ \ ,\nen
\{Q_\alpha,Q_\beta\}&\seq& i (\gamma_a)_{\alpha\beta} J^a\ \ .\eea
As these equations suggest, $J_a(Q_\alpha)$ are Grassmann even(odd) generators
with $\alpha=1,2$ and the $J_a$ forming a $SU(2)$ subgroup.  The $(\gamma^a)$
are three dimensional Clifford matrices:
$\gamma^a\in(\sigma^3,-\sigma^1,\sigma^2)$.
\newcommand{\IA}{{\slash \kern -0.4em A}}
Introduce a connection one-form on the $OSp(1|2)$-bundle over $X$,
\be
{\tt A}\seq \ca^aJ_a\pls \Upsilon^\alpha Q_\alpha\ \ ,\ee
where the two component Grassmann odd field $\Upsilon^\alpha$ is
$(\psi^1,\psi^2)$.  Its curvature two-form is
\bea
\cf&\seq& {\hat F}^a J_a\pls f^\alpha Q_\alpha\ \ ,\nen
{\hat F}^a&\seq& F^a(\ca)J_a \pls i\half(\gamma^a)_{\alpha\beta}
\Upsilon^\alpha\wedge \Upsilon^\beta\, J_a\ \ ,\nen
f^\alpha&\seq& D(\ca)\Upsilon^\alpha\seq d\Upsilon^\alpha \mi\half
\ca^a\wedge\Upsilon^\beta(\gamma_a)_\beta{}^\alpha\ \ ,\eea
where $F(\ca)$ is the usual curvature of a $SU(2)$-bundle.  It then follows
that the self-dual curvature equations become
\be
\cf^+\seq0\qquad \Longrightarrow\qquad\left\{
\begin{array}{lc}
F^{+\, a}(\ca)\seq \mi i\half (\gamma^a)_{\alpha\beta}P_+(\Upsilon^\alpha\wedge
\Upsilon^\beta) \ ,\\
P_+(D(\ca)\Upsilon^\alpha)\seq 0\ \ .
\end{array}
\right.
\ee
By enlarging the principal bundle we have incorporated the monopole equations
into a single vanishing self-dual curvature equation.

\section{Conclusion}\label{S_Conc}

\forcepar
We have realized the twisted version of the monopole equations \cite{W(mon)} in
the fields space of topological Yang-Mills.  In addition, we have identified a
unitary transformation between the respective cohomologies and were led to
conjecture an equivalence between the Donaldson invariants and those which
follow from the spectral flows governed by the  monopole equations.
Our method suggests a number of generalizations, including the appearance of
the monopole equations in a self-dual curvature condition on a super-bundle. In
addition, we have found that the twisted monopole equations arise directly from
vanishing self-dual curvature condition on an $SU(2)$-bundle in which the
fermions are realized as spanning the tangent space for a specific background.
That background being one in which only the connection in the abelian direction
is non-zero and its tangent space is null.  In this vein, the fermions in the
equations  span the tangent space to the point zero which is the background
value of the connections in the compliment of the $U(1)$ subgroup.

\pagebreak


\begin{thebibliography}{99}

\bibitem{W(mon)}{E. Witten, ``Monopoles and Four-Manifolds", hep-th/9411102}

\bibitem{W(TYM)}{E. Witten, \cmp{117} (1988) 353}

\bibitem{Fl}{A. Floer, {\it Bull. Am. Math. Soc.} {\bf 16} (1987) 279;
\cmp{118} (1988) 215}

\bibitem{Don}{S. K. Donaldson, {\it Topology} {\bf 29} (1990) 257}

\bibitem{W(SYM)}{E. Witten,  {\it J. Math. Phys.} {\bf 35} (1994) 5101}

\bibitem{SW}{N. Seiberg and E. Witten, \np{426} (1994) 19; ``Monopoles, Duality
and Chiral Symmetry Breaking in N=2 Supersymmetric QCD," hep-th/9408099}

\bibitem{KM1}{P. Kronheimer and T. Mrowka, {\it Bull. Am. Math. Soc.} {\bf 30}
(1994) 215}

\bibitem{KM2}{P. Kronheimer and T. Mrowka, ``The genus of embedded surfaces in
the projective plane," {\it Math. Res. Letters}, to apppear}

\end{thebibliography}
\end{document}